\documentclass{aastex}

\usepackage{spr-astr-addons}
\usepackage{url}
\usepackage{natbib}

\RequirePackage{color}

\begin{document}

\title{Mass evaluation for red giant stars based on the 
spectroscopically determined atmospheric parameters}
\shorttitle{Mass evaluation for red giants}
\shortauthors{Y. Takeda}

\author{Yoichi Takeda\altaffilmark{1}}
\altaffiltext{1}{11-2 Enomachi, Naka-ku, Hiroshima-shi 730-0851, Japan\\
E-mail: ytakeda@js2.so-net.ne.jp} 

\begin{abstract}
The mass ($M$) of a star can be evaluated from its spectroscopically determined   
effective temperature ($T_{\rm eff}$) and metallicity ([Fe/H]) along with the 
luminosity ($L$; derived from parallax), while comparing them with grids of 
theoretical evolutionary tracks. It has been argued, however, that such a 
track-based mass ($M_{\rm trk}$) may tend to be overestimated for the case 
of red giants.  
Meanwhile, there is an alternative approach of evaluating mass ($M_{gLT}$)
directly from surface gravity ($g$), $L$, and $T_{\rm eff}$.   
The practical reliability of $M_{gLT}$ was examined for $\sim 100$ benchmark giants 
in the {\it Kepler} field, for which atmospheric parameters are already determined 
and the reliable mass ($M_{\rm seis}$) along with the evolutionary status are 
known from asteroseismology. In addition, similar check was also made for the 
accuracy of $M_{\rm trk}$ for comparison.
It turned out that, while a reasonable correlation is seen between $M_{gLT}$ 
and $M_{\rm seis}$ almost irrespective of the stellar property, its precision is rather
insufficient because $\log (M_{gLT}/M_{\rm seis})$ distributes rather widely within 
$\sim \pm$0.2--0.3~dex. 
In contrast, the reliability of $M_{\rm trk}$ was found to depend on the evolutionary
status. Although $M_{\rm trk}$ and $M_{\rm seis}$ are satisfactorily consistent with 
each other (typical dispersion of $\log (M_{\rm trk}/M_{\rm seis})$ is within 
$\sim \pm$~0.1~dex) for H-burning red giants as well as He-burning 2nd clump giants of 
higher mass, $M_{\rm trk}$ tends to be considerably overestimated as compared to 
$M_{\rm seis}$ by up to $\la$~0.4~dex for He-burning 1st clump giants of lower mass. 
Accordingly, $M_{gLT}$ and $M_{\rm trk}$ are complementary with each other in terms of 
their characteristic merit and demerit. 
\end{abstract}


\section{Introduction}

\subsection{Mass determination with evolutionary tracks}

The mass ($M$) of a star is of decisive importance in astrophysics, because it 
essentially governs the stellar properties and their variations during the lifetime.
The conventional method for determining this key parameter is to compare the position 
on the theoretical Hertzsprung--Russell (HR) diagram, where the bolometric
luminosity of a star ($L$; determinable from the observed brightness
along with the distance) is plotted against its effective temperature ($T_{\rm eff}$),
with a set of stellar evolutionary tracks.
Since these theoretical tracks (loci on the $L$ vs. $T_{\rm eff}$ plane) are computed 
for various combinations of $M$ and metallicity ($z$; often regarded as equivalent 
to the Fe abundance [Fe/H] in practice), $M$ can be reasonably determined
by finding the best-match between ($L$, $T_{\rm eff}$)$_{\rm obs}$, and
($L$, $T_{\rm eff}$)$^{\{M,z\}}_{\rm cal}$, if $L$, $T_{\rm eff}$, and [Fe/H] ($z$)
are given as input observational parameters. Such established $M$ is referred to as 
$M_{\rm trk}$ (track-based mass) in this paper.

This mass-evaluation technique is occasionally called ``spectroscopic method''(despite 
that it is essentially based on theoretical stellar evolution calculations), because 
the use of spectroscopically determined $T_{\rm eff}$ and [Fe/H] is almost prerequisite 
(since sufficient precision can not be attained by other simpler method such as the 
photometric one, especially for [Fe/H]). The widely adopted approach to accomplish 
this spectroscopic pre-conditioning is to make use of a number of Fe~{\sc i} and 
Fe~{\sc ii} lines, by which $T_{\rm eff}$, $\log g$ (surface gravity), $v_{\rm t}$ 
(microturbulence), and [Fe/H] can be effectively established by the requirements of 
(i) excitation equilibrium of Fe~{\sc i} (abundances do not depend upon the excitation 
potential), (ii) ionisation equilibrium between Fe~{\sc i} and Fe~{\sc ii} (equality
of the mean abundances), and curve-of-growth matching for Fe~{\sc i} (abundances do 
not depend upon the line strength) (see, e.g., Takeda, Ohkubo \& Sadakane, 2002).

\subsection{Suspicion on $M_{\rm trk}$ for red giants}

However, the credibility of $M_{\rm trk}$ evaluated in this way for red giants (evolved 
stars with typical masses of $\sim$~1.5--3~$M_{\odot}$) has become a controversial issue.
The background is that the importance of reliable mass determination for such cool giant 
stars has grown these days, especially with regard to 
the research field of extrasolar planets. Namely, since searching planets for such 
intermediate-mass stars is not easy when they are hot stars on the upper main sequence, 
those in the evolved red-giant stage (sometimes called ``retired A-type stars''
as done by Johnson et al., 2007) are 
more suitable and thus mainly targeted. As such, knowing the mass of the host giant is 
of critical importance, 
because it is one of the key parameters (like metallicity) 
affecting the physical properties and formation process of planets 
(e.g., Ghezzi, Montet, \& Johnson, 2018; Wolthoff et al., 2022).

The doubt for the reliability of $M_{\rm trk}$ emerged almost a decade ago, when the novel 
approach of precise mass determination for red giants became possible by using the 
asteroseismic method based on very high-precision photometric observations from satellites 
(CoRoT, {\it Kepler}, TESS). That is, the track-based $M_{\rm trk}$ tends to be systematically 
larger compared to the seismologically established mass ($M_{\rm seis}$), which suggests that 
the former would have been more or less overestimated. As a matter of fact, this 
possible overestimation of $M_{\rm trk}$ 
has been reported and argued in quite a few number of studies; e.g., 
Lloyd (2011, 2013); Schlaufman \& Winn (2013); Johnson et al. (2014); Sousa et al. (2015);
Ghezzi \& Johnson (2015); Takeda \& Tajitsu (2015); Takeda et al. (2016);
Hj{\o}rringgaard et al. (2017); Stello et al. (2017); North et al. (2017); 
Campante et al. (2017); White et al. (2018); Ghezzi, Montet, \& Johnson (2018); 
Malla et al. (2020); Wolthoff et al. (2022).

It is not clear, however, $M_{\rm trk}$ results for red giants of which type are 
seriously affected by errors or they may alternatively be regarded as practically usable, 
because the conclusions in these papers are rather diversified (some authors state that
spectroscopic $M_{\rm trk}$ determination using grids of evolutionary tracks is not 
significantly affected by systematic errors and thus valid; e.g., Ghezzi \& Johnson, 2015).
In any event, we should be aware of the possibility that $M_{\rm trk}$ derived 
for giant stars may suffer appreciable errors, owing to the increasing difficulty
of its application to red giants due to closely-lying (or even crossing) evolutionary 
tracks of different masses.

\subsection{Mass derivation from surface gravity}

Incidentally, there is another way of mass determination, which may also be called 
as ``spectroscopic method'' (like the case of $M_{\rm trk}$) because  
it is similarly based on the spectroscopically established atmospheric parameters
($T_{\rm eff}$ and $\log g$) along with $L$ (from the parallax).
The point is to make use of the information of $M$ contained in the surface gravity ($g$), 
That is, combining the two relations, $g \propto M/R^{2}$ ($R$: radius) and 
$L \propto T_{\rm eff}^{4} R^{2}$, we obtain $M \propto g L /T_{\rm eff}^{4}$, or
\begin{equation}
\begin{split}
 \log (M/M_{\odot}) & = \log (g/g_{\odot}) \\
                    &+ \log (L/L_{\odot}) - 4 \log (T_{\rm eff}/T_{{\rm eff}\odot}).
\end{split}
\end{equation}
The mass determined from $g$, $L$, and $T_{\rm eff}$ by Eq.~(1) is hereinafter referred
to as $M_{gLT}$. Although this is a simple and straightforward approach without 
any necessity of using theoretical models (evolutionary tracks), it seems to have barely 
been employed in practice to the author's knowledge for giants (though
sometimes discussed for the case of dwarfs; e.g., Valenti, \& Fischer, 2005). 
This is presumably because 
accomplishing sufficient precision in spectroscopic $\log g$ has been considered as 
rather difficult, especially for giant stars of lower density atmospheres.
Actually, comparisons of the spectroscopically determined $\log g$ values for 
red giants in various literature suggest that they show an appreciable diversity 
(see, e.g., Fig.~5b, Fig.~6b, Fig.~7b, and Fig.~8b in Takeda, Sato \& Murata, 2008;
hereinafter referred to as T08),
which indicates that $\log g$ determination is not robust but depends on technical details. 
   
However, the author has come to think that spectroscopic $\log g$ may be 
relied upon based on the results of Takeda \& Tajitsu (2015; hereinafter called as T15) 
and Takeda et al. (2016: T16), in which spectroscopically determined $\log g$ values of 
$\sim 100$ red giants in the {\it Kepler} field and those established by the 
seismological technique were compared and found to be in satisfactory agreement with 
each other (see Fig.~A1 in T16).\footnote{
In the case of FGK-type main sequence stars, Brewer et al. (2015) arrived at a similar 
conclusion that their results of spectroscopic $\log g$ are in reasonable consistency 
with seismic gravities. 
} 
Therefore, considering that the parallax data necessary for evaluation of $L$ (many of 
which were unknown at the time of T15 and T16) are now available for all of the sample stars 
thanks to the {\it Gaia} DR2 database (Gaia Collaboration et al., 2016, 2018), it is interesting and
worthwhile to examine whether and how the $M_{gLT}$ values of these {\it Kepler} giants (derived 
from Eq.~(1) by using the atmospheric parameters already determined spectroscopically 
in T15 and T16) are compared with the seismic ones ($M_{\rm seis}$). 
This is the main purpose of this study.

Besides, by taking this opportunity, the problem on the reliability of $M_{\rm trk}$ 
is revisited for these $\sim 100$ red giant stars. Since Ghezzi \& Johnson (2015) concluded 
that $M_{\rm trk}$ values of their 59 benchmark giant stars derived by the open software PARAM\footnote{
http://stev.oapd.inaf.it/cgi-bin/param\_1.3/
} 
(da Silva et al., 2006), which is based on the Bayesian estimation method, are reasonably
consistent with the reference mass values (derived either from orbital motions of binaries 
or by the seismic method), we likewise make use of the PARAM code to evaluate $M_{\rm trk}$
for our sample stars in the {\it Kepler} field, for which not only the mass but also the 
evolutionary status is seismologically established.  Do we obtain a consistency between 
$M_{\rm trk}$ and $M_{\rm seis}$? Is there any dependence upon the evolutionary stage? 
To check these points is the second aim of this investigation.   
   
\section{Program stars}

The main targets in this investigation are the 103 red giants in the {\it Kepler} 
field studied in T15 and T16 based on the high-dispersion spectra, most\footnote{
The spectra for 13 (out of 103) stars were taken from Thygesen et al. (2012).
Regarding 3 stars (KIC~1726211, KIC~2714397, KIC~3744043) for which both Subaru data
and Thygesen et al.'s (2002) data are available, the results based on the former
Subaru spectra are adopted in this study.    
} of which were obtained on 2014 September 9 and 2015 July 3 (UT) by 
using the High-Dispersion Spectrograph (HDS) of the Subaru Telescope.  
Their atmospheric parameters ($T_{\rm eff}$, $\log g$, [Fe/H], and $v_{\rm t}$) were 
determined from the equivalent widths of Fe~{\sc i} and Fe~{\sc ii} lines 
by using the TGVIT\footnote{
https://www2.nao.ac.jp/\~{}takedayi/tgv/
} program (Takeda et al., 2005), which is the update
version of the TGV code (Takeda et al., 2002). The seismic mass values ($M_{\rm seis}$; 
to be used as the comparison reference) are the same as those adopted in T15 and T16, 
which were derived from the spectroscopic $T_{\rm eff}$ along with the seismological 
frequencies ($\nu_{\rm max}$ and $\Delta \nu$; taken from Mosser et al., 2012)  
according to Eq.~(2) of T15. Likewise, the evolutionary status (RG or RC1 or RC2;
cf. the caption of Fig.~3) of each star is already determined by Mosser et al. (2012)

In addition, for the purpose of comparison with Ghezzi \& Johnson's (2015) results,
11 nearby field giants were also included, which are common to 
their 59 benchmark stars (cf. their Table~1) and whose atmospheric parameters are 
already determined in the previous papers of the author by using the TGVIT program. 
The $M_{\rm seis}$ values of 9 stars were calculated as in T15 by using $\nu_{\rm max}$ 
and $\Delta \nu$ given in their Table~1, while (exceptionally) the reference masses 
of the components of Capella system (cooler primary and hotter secondary) 
are the results of Torres et al.'s (2015) orbital motion analysis.
A total of 114 (=103+11) program stars and their stellar parameters are presented
in Table~1.

\section{Evaluation of luminosity}

The luminosity $L$, another necessary parameter along with $T_{\rm eff}$
and $\log g$ in Eq.(1), is related with $V$ (apparent visual magnitude in mag), 
$\pi$ (parallax in milliarcsecond) by the following equation
\begin{equation}
\begin{split}
 -2.5 \log (L/L_{\odot}) & = V + 5 + \log(\pi/1000) \\
                         & - A_{V} + {\rm BC} - 4.75,
\end{split}
\end{equation}
where $A_{V}$ is the interstellar extinction (in mag), BC is the bolometric correction
(in mag), and 4.75 (mag) is the bolometric absolute magnitude of the Sun.

The $V$ data of the program stars were obtained by consulting the SIMBAD database
wherever possible. However, since the relevant data were unavailable for 
13 {\it Kepler} giants, their apparent visual magnitudes were tentatively derived
from the $r$-band magnitude ($r^{\rm KIC}$) given in the {\it Kepler} Input Catalogue 
(Brown et al. 2011; hereinafter abbreviated as KIC) by the following procedure. 
\begin{itemize}
\item
First, for each {\it Kepler} giant with known $V_{\rm simbad}$, theoretical $(V-R)_{\rm cal}$ 
colour index corresponding to $T_{\rm eff}$ and $\log g$ is evaluated by using 
Girardi et al.'s (2000) grids of stellar evolution calculations.  
\item
Then, this $(V-R)_{\rm cal}$ is applied to $r_{\rm KIC}$ to derive $v$ 
by the relation $v = (r^{\rm KIC} - A_{r}^{\rm KIC})+(V-R)_{\rm cal} + A_{V}^{\rm KIC}$,  
where $A_{V}^{\rm KIC}$ is the interstellar extinction given in KIC
and $A_{r}^{\rm KIC} = 0.88 A_{V}^{\rm KIC}$ (cf. Brown et al., 2011).
\item
Such obtained $v$ is plotted against $V_{\rm simbad}$ in Fig.~1a,
which shows that the relation between these two is can be approximately expressed
as $v = 0.934 V + 1.002$.
\item
Accordingly, since $v'$ (corrected $v$) defined by the relation
$v' \equiv (v - 1.002)/0.934$ is nearly equivalent to $V$ (almost
free from systematic deviation; cf. Fig.~1b), this $v'$ is adopted
as $V$ for each of the 13 {\it Kepler} giants for which $V$ data
are not available in SIMBAD.
\end{itemize}

The parallaxes ($\pi$) of all 114 program stars were taken from the 
SIMBAD database, which are mostly from {\it Gaia} DR2 and of 
sufficient precision (errors are a few percent or less).\footnote{
Several recent studies have reported that small corrections should be applied to the 
original {\it Gaia} parallaxes. This effect is separately discussed in the Appendix.
}

The bolometric correction (BC) for each star was evaluated from $T_{\rm eff}$ and [Fe/H] 
by using  Eq.(17) and Eq.(18) (both results were averaged at the overlapping $T_{\rm eff}$ range)
of Alonso, Arribas, \& Mart\'{\i}nez-Roger (1999).

Since the relevant red giants in the {\it Kepler} field are rather distant (distances 
up to $\sim 2$~kpc) and of low galactic latitude (typically $b \sim 10$~deg), they suffer
appreciable interstellar extinction ($A_{V}$), for which two data sources were examined:  
(i) application of the EXTINCT\footnote{http://asterisk.apod.com/library/ASCL/extinct/extinct.for}
 program (Hakkila et al., 1997) which yields averaged $A_{V}$ 
and its error (based on various different extinction studies) if the data of distance 
and galactic coordinates are given, and (ii) $A_{V}$ data given in KIC.
These two kinds of $A_{V}$ and the difference between them are plotted against the distance 
in Figs.~2a--2c.
While we can see from these figures that both are mostly consistent with each other,
$A_{V}^{\rm EXTINCT}$ is prominently larger than $A_{V}^{\rm KIC}$ for 10 stars  
deviating from the mean trend (cf. Fig.~2a). However, these anomalous $A_{V}^{\rm EXTINCT}$
values are questionable, because the corresponding $L$ values obtained by Eq.(2) tend to become more 
discrepant in comparison with $L_{\rm seis}$ derived in T16 (from spectroscopic $T_{\rm eff}$ and 
seismic $R$) as shown in Figs.~2d and 2e. Therefore, we decided to adopt $A_{V}^{\rm KIC}$ 
wherever possible. $A_{V}^{\rm EXTINCT}$ values were employed only for 9 {\it Kepler} giants 
(for which $A_{V}^{\rm KIC}$ are not available) and for the additional 11 nearby giants.  

The adopted data of $V$, $A_{V}$, $\pi$, and BC along with the resulting $\log L$  
from Eq.(2) are summarised in Table~1, where the $M_{gLT}$ values derived from Eq.(1)
are also given. Besides, more complete data (including errors and data sources for 
$V$ or $A_{V}$) are presented in the electronic data table (``stellar\_parameters.dat'')
available as the online material.

Now that $T_{\rm eff}$ and $L$ have been established for all of the 114 stars, 
their locations in the $\log L$ vs. $\log T_{\rm eff}$ diagram are plotted in Fig.~3
(RG, RC1, and RC2 are discriminated by different symbols in the same manner as in T16),  
where the theoretical PARSEC tracks (Bressan et al., 2012, 2013) corresponding to 
$z = 0.01$ are also drawn for comparison.

\section{Mass from evolutionary tracks}

In addition to such derived $M_{gLT}$, the track-based mass ($M_{\rm trk}$) was 
also determined for all the sample stars. Since the main motivation for doing this was to check 
Ghezzi \& Johnson's (2015) conclusion supporting the practical validity of $M_{\rm trk}$ 
(without any significant errors) derived for red giants, the open software tool PARAM 
(da Silva et al. 2006) was employed as they did, which determines the best stellar parameters 
and their errors by comparing the position on the HR diagram with grids of evolutionary 
tracks based on the Bayesian estimation method.
We used the version 1.3 of PARAM, which returns $age$, $M_{\rm trk}$, $\log g$, $R$, and 
$(B-V)_{0}$ along with their errors as output, for given 8 input parameters (observed 
values and errors of $T_{\rm eff}$, [Fe/H], $V_{0} (\equiv V - A_{V})$, and $\pi$.
As to the errors of input parameters, while the actual error attached to each value was 
assigned for $\pi$, typical values of $\pm 100$~K and $\pm 0.1$~dex were uniformly 
assumed for $T_{\rm eff}$ and [Fe/H], respectively (cf. Sect.~3.3 in T08). 
Regarding $V_{0}$, since the main source of error is due to the uncertainty in $A_{V}$,
$\pm 0.3$~dex (103 {\it Kepler} giants; judged from the size of error bars in Fig.~2a) 
and $\pm 0.1$~dex (11 nearby giants) were given.\footnote{
The error of $A_{V}$(EXTINCT) is evaluated as the standard deviation of the mean 
of those derived by different subroutines (taken from various literature 
using different dust maps), as described in the header of the EXTINCT program.
Such obtained error in $A_{V}$(EXTINCT) was further assumed to be equal to 
that of $A_{V}$(KIC), because there is no other way for estimating $A_{V}$(KIC) errors.
Assigning $\pm 0.1$~mag to $V$ errors of nearby giants is just tentative, owing to the 
lack of information on errors in absolute photometry for such bright stars, 
though this may be a somewhat overestimation.
}

The resulting values of $M_{\rm trk}$ are presented in Table~1, while the complete input data 
and output results of PARAM\footnote{
In implementing PARAM1.3, the option of PARSEC version 1.1 (Bressan et al., 2012) was selected
for the evolutionary tracks, and the default setting was applied unchanged for the Bayesian priors.
} are summarised in ``PARAMresults.dat'' of the online material.  
The $M_{\rm trk}$ values of 11 nearby stars derived in this study are compared with 
those of Ghezzi \& Johnson (2015: cf. their Table~2) in Fig.~4. It can be seen from this
figure that both determinations are almost consistent with each other, except for
HD~185351 which shows some discrepancy due to the difference in [Fe/H] (i.e., they used 
+0.16 in contrast to the adopted 0.00 derived in T08). 

\section{Discussion}

\subsection{$M_{gLT}$ versus $M_{\rm trk}$}

We are now ready to discuss the quantitative reliability of $M_{gLT}$ (derived by the 
straightforward model-independent approach) as well as of $M_{\rm trk}$ (determined 
with the help of theoretical evolutionary tracks) by comparing them with the reference 
$M_{\rm seis}$ values. In Fig.~5, these two kinds of stellar masses are plotted against 
$M_{\rm seis}$, and the dependences of relative differences (logarithmic mass ratios) 
upon $M_{\rm seis}$, [Fe/H], and $\log L$ are also depicted, from which several 
characteristic trends of interest can be read.

Fig.~5a suggests that $M_{gLT}$ correlates with $M_{\rm seis}$, but the correlation is 
not so tight with a rather large dispersion (correlation coefficient is +0.62).
Actually, the logarithmic mass ratio distributes at $-0.3 \la \log (M_{gLT}/M_{\rm seis}) \la +0.3$
(as shown in Figs.~5b--5d) and its average is $\langle \log (M_{gLT}/M_{\rm seis}) \rangle = +0.05$ 
(standard deviation is $\sigma = 0.14$). Accordingly, $M_{gLT}$ by itself is not so sufficiently
precise as to be practically usable with confidence.
Meanwhile, since the disagreement of $M_{gLT}$ relative to $M_{\rm seis}$ almost
uniformly distribute without a significant dependence upon the stellar parameters 
(except that $\log (M_{gLT}/M_{\rm seis})$ tends to increase with a decrease in mass 
at $M_{\rm seis} \la 1$~$M_{\odot}$; cf. Fig.~5b), it would be meaningful to discuss 
$M_{gLT}$ values for a large number of stars altogether because errors may be statistically 
cancelled, which may be regarded as a point of merit.

On the other hand, $M_{\rm trk}$ shows opposite characteristics in the sense that
whether it is consistent with $M_{\rm seis}$ apparently depends upon the evolutionary stage
as manifestly seen in Figs.~5e--5h. That is, while a satisfactory agreement is observed in stars of 
RG (H-burning) and RC2 (He-burning, mass higher than 1.8~$M_{\odot}$), $M_{\rm trk}$ 
is systematically larger than $M_{\rm seis}$ for the case of RC1 (He burning, mass lower 
than 1.8~$M_{\odot}$).\footnote{
Accordingly, this consequence (considerable $M_{\rm trk} > M_{\rm seis}$ 
discrepancy is seen for lower-mass RC1 giants of $M < 1.8 M_{\odot}$) 
apparently contradicts the result of Malla et al. (2020) who argued based on
a sample of 16 giants that large overestimation of spectroscopic masses 
occurs for higher mass giants of $M > 1.6 M_{\odot}$).
} 
As a matter of fact, $\log (M_{\rm trk,RC1}/M_{\rm seis})$ significantly
deviates from zero extending up to $\sim +0.4$ 
($\langle \log (M_{\rm trk,RC1}/M_{\rm seis}) \rangle = +0.16$ with $\sigma = 0.11$),
while the agreement with $M_{\rm seis}$ is fairly good for both $M_{\rm trk,RG}$ and $M_{\rm trk,RC2}$ 
($\langle \log (M_{\rm trk,RG}/M_{\rm seis}) \rangle = -0.01$ with $\sigma = 0.06$ and 
$\langle \log (M_{\rm trk,RC2}/M_{\rm seis}) \rangle = -0.01$ with $\sigma = 0.08$).
More precisely, those RC1 stars showing conspicuous deviations are confined in the luminosity
range of $1.7 \la \log (L/L_{\odot}) \la 2.0$ (cf. Fig.~5h), 
and their masses are comparatively low in the range of 
$0.8 \la M_{\rm seis}/M_{\odot} \la 1.5$ (cf. Fig.~5f). 
The reason for this problem is presumably attributed to the fact that RC1 tracks of 
lower mass (e.g., $\sim 1$~$M_{\odot}$) and RG/RC2 tracks of higher mass (e.g., 
$\sim 2$~$M_{\odot}$) are intricate with each other at this luminosity range (cf. Fig.~3).
The double peak in the mass probability function inevitably caused by the degeneracy 
of tracks would eventually result in an erroneous $M_{\rm trk}$ (though it is not
straightforward to explain why overestimation of RC1 mass is preferentially observed 
but underestimation of RG/RC2 mass is not). 
Accordingly, Ghezzi \& Johnson's (2015) argument supporting the validity of $M_{\rm trk}$ 
determined for red giants with the help of evolutionary tracks is not necessarily correct, 
because $M_{\rm trk}$ is apt to be considerably overestimated for lower-mass RC1 stars at 
$1.7 \la \log (L/L_{\odot}) \la 2.0$, even though sufficient precision is surely 
attained (as they say) for stars of RG and RC2 stages.
The reason why Ghezzi \& Johnson (2015) did not detect appreciably large overestimation of 
$M_{\rm trk}$ (in comparison with $M_{\rm seis}$) may be due to the difference in the mass 
distribution of sample stars. That is, since their program stars include only a small 
number of stars in the lower-mass range around $\sim 1$~$M_{\odot}$ (where RC1 giants 
show large mass discrepancies as revealed in this study) as seen from their Fig.~2,
this problem may not have been particularly noticeable.

Combining what has been described above, we can summarise as follows regarding the 
merit and demerit of $M_{gLT}$ and $M_{\rm trk}$ and how to make the best use of them.
\begin{itemize}
\item
The precision of $M_{gLT}$ itself is unfortunately insufficient for practical use, because
errors on the order of several tens percent or more ($\sim 0.2$~dex) are common.
It is not recommendable to put much trust in its value.  
\item
In contrast, $M_{\rm trk}$ is comparatively more reliable because its involved errors can be 
apparently smaller (typically within $\la$10--20\% for the case of RG/RC2) but its weakpoint 
is that a considerable overestimation (even by $\ga 100$\% in unfortunate cases) takes place
 for a specific group (RC1) of stars.   
\item
Therefore, since the evolutionary status of red giants is generally unknown, the hybrid 
approach utilising both together would be advisable, where larger weight may be placed 
to $M_{\rm trk}$ while $M_{gLT}$ plays the checking role. Actually, spectroscopically 
determined atmospheric parameters ($T_{\rm eff}$, $\log g$, and [Fe/H])
can be most effectively exploited by calculating both $M_{gLT}$ (derived from $L$, $\log g$, 
and $T_{\rm eff}$) and $M_{\rm trk}$ (determinable from $L$, [Fe/H], and $T_{\rm eff}$).
\item
Practically, $M_{\rm trk}$ would be allowed to adopt at first, though it should not be
blindly trusted. Especially, one should be cautious for stars in the luminosity range of 
$1.7 \la \log (L/L_{\odot}) \la 2.0$. In such cases, it would be a useful 
check to compare $M_{\rm trk}$ and $M_{gLT}$ with each other in order to avoid making 
serious errors. 
\item
Although $M_{gLT}$ does not have a sufficient precision by itself, this rather large 
uncertainty applies similarly to any star irrespective of the stellar types or parameters. 
As such, studying the general trends of $M_{gLT}$ based on a large sample of stars may
yield almost correct results, because random errors would be cancelled in such a statistical 
discussion.
\end{itemize}

\subsection{Age--metallicity relation revisited with $M_{gLT}$}

As an example of applying $M_{gLT}$ to discussing the trends of many stars, 
the mass-dependent parameters of 322 mid-G to early-K giants studied in T08 were
reinvestigated by using $M_{gLT}$.
Actually, there was a problem in the consequence of that paper. Although $M_{\rm trk}$ was 
determined in T08 for each star by using Lejeune \& Schaerer's (2001) grids of evolutionary tracks,
the resulting stellar age (closely associated with the mass) yielded an age--metallicity 
relation for giants seriously discrepant from that of dwarfs (cf. Fig.~14 therein).
T15 revisited this problem and concluded that $M_{\rm trk}$ results of
many (even if not all) clump giants determined in T08 are likely to be significantly
overestimated (see also Appendix~B in T16), leading to an underestimation of corresponding ages. 
It is thus interesting to see how the results would change if $M_{gLT}$ is used instead
of $M_{\rm trk,T08}$.
 
The values of $M_{gLT}$ were determined for each of the 322 giants by using $T_{\rm eff}$, 
$\log g$, and $\log L$ given in Table~1 of T08. The comparison between such obtained $M_{gLT}$
and $M_{\rm trk,T08}$ is shown in Fig.~6a. Besides, how $T_{\rm eff}$, $\log g$, [Fe/H],
and $\log L$ are related with $M_{gLT}$ is depicted in Figs.~6b, 6c, 6d, and 6e, respectively 
(which should be compared with Figs.~3d, 3e, 3f, and 3a in T08).
Further, the age of each star was derived from $M_{gLT}$ by applying the approximate relation 
$\log age \simeq 10.74 - 1.04 (M/M_{\odot}) + 0.0999 (M/M_{\odot})^{2}$
(where $age$ is in yr; cf. Sect.~3.2 in T08), and the resulting [Fe/H] vs. $age$ relation
is displayed (along with that of dwarfs) in Fig.~6f.

Fig.~6a reveals that an inequality relation $M_{gLT} < M_{\rm trk,T08}$ holds at 
$M_{gLT} \la 3$~$M_{\odot}$, which indicates that $M_{\rm trk,T08}$ was systematically 
overestimated in this mass range. The  $age$ derived by using $M_{gLT}$ 
(instead of $M_{\rm trk,T08}$) results in a revised [Fe/H] vs. $age$ relation of giants,
which is more consistent and well overlaps with that of dwarfs (Fig.~6f).
This reasonably confirms the argument done in T15 (cf. Fig.~13 therein) that the discrepancy 
between giants and dwarfs would be effectively mitigated by the revision of $M$.       

\subsection{On the error source of $M_{gLT}$}

As described in Sect.~5.1, the error involved in $M_{gLT}$ turned out unfortunately somewhat 
too large for the purpose of practical application.  
On the assumption of random errors in the Gaussian distribution , $\sigma(\log M_{gLT})$ 
is written as the root-sum-square of $\sigma(\log g)$, $\sigma(\log L)$, and 
$\sigma(4\log T_{\rm eff})$ according to Eq.(1). In the present case of red giants in the
{\it Kepler} field, $\sigma(\log g) \sim 0.1$ (cf. Table~3 in T15), 
$\sigma(\log L) \sim 0.1$ (mainly determined by errors in $A_{V}$ on the order of $\sim 0.3$~mag;
cf. Sect.~4), and $\sigma(4\log T_{\rm eff}) \sim 0.04$ (corresponding to $\sim 100$~K).
These values yield $\sigma(\log M_{gLT}) \sim \sqrt{0.1^{2}+0.1^{2}+0.04^{2}} \sim 0.15$,
which is reasonably consistent with the standard deviation (0.14~dex) observed for the 
distribution of $\log (M_{gLT}/M_{\rm seis})$. 

Among these error terms, while $\sigma(\log L)$ may be brought down to the level of 
$\sim 0.05$~dex (if $A_{V}$ is insignificant) and the contribution of $\sigma(4\log T_{\rm eff})$ 
is already small, $\sigma(\log g)$ is most important (essentially controlling
the accuracy of $M_{gLT}$) but its further reduction would not be easy. 
On the contrary, this error may become more serious, because spectroscopic $\log g$ 
determination tends to sensitively depend upon technical details.
Actually, the accuracy of $\log g$ appears to differ from case to case 
as seen from the comparison of various literature values (see Sect.~3.3 in T08). 
Therefore, the precision of $\sigma(\log g) \simeq 0.1$ accomplished in this investigation
may not necessarily be guaranteed in other circumstances. 
  
As an example to illustrate this situation, the atmospheric parameters spectroscopically determined 
by Luck (2015) based on Fe~{\sc i} and Fe~{\sc ii} lines in his extensive study of 1133 FGK-type 
giants are compared with those of T08 in Fig.~7. It can be seen from Fig.~7b that the comparison
between $\log g$(Luck) and $\log g$(T08) shows an appreciable systematic difference with a 
rather large dispersion as $\langle \Delta \log g \rangle$(Luck$-$T08) = $-0.18$ ($\sigma = 0.19$),\footnote{
Although the cause of this discrepancy is not clear, the difference in the adopted set of Fe~{\sc i} 
and Fe~{\sc ii} lines might be involved. The number of lines used by Luck (2015) amounts
up to $\sim$~500--600 (Fe~{\sc i}) and $\sim$~50--60 (Fe~{\sc ii}). These numbers are 
considerably larger (by a factor of $\sim$~2--3) compared to the case of the author's previous 
studies (T08, T09, T15, T16, T19), where the adopted lines were limited to those weaker 
than 120~m\AA, because stronger ones affected by the effect of damping wings were intentionally 
avoided (as they tend to be problematic in view of the difficulty in equivalent-width 
measurements or uncertainties in damping parameters). 
Accordingly, if such stronger lines were included in Luck's (2015) analysis,
this might be the reason for the disagreement.  
} despite that reasonable consistency is observed in other spectroscopic parameters (Figs.~7a, 7c,
and 7d). 
As such, using $\log g$ derived by Luck (2015) would result in systematic underestimation 
of $M_{gLT}$ (on the average by $\sim 0.2$~dex or $\sim 60$\% along with 
an appreciably larger random dispersion) compared to the case based on $\log g$(T08).

Accordingly, attention should be paid to the precision of $\log g$ in case of deriving 
$M_{gLT}$, for which confirming its quantitative reliability is prerequisite. It may be  
useful to carry out a test in advance on a number of benchmark giant stars with known 
seismic parameters (e.g., by using the spectra of {\it Kepler} giants published by 
Thygesen et al., 2012), in order to check the consistency between spectroscopic and 
seismic gravities. 
 
\section{Summary and conclusion}

The conventional method for determining the mass of a star is to compare its location 
on the $\log L$ vs. $T_{\rm eff}$ diagram with grids of theoretical evolutionary tracks 
computed for various combinations of ($M$, [Fe/H]). This is occasionally called
``spectroscopic mass determination'' because  $T_{\rm eff}$ and [Fe/H] are to be 
established in advance (in addition to $L$ determinable from the observed magnitude and 
the parallax) by the spectroscopic approach using Fe~{\sc i} and Fe~{\sc ii} lines.

Recently, the validity of such track-based mass ($M_{\rm trk}$) determined for red giant 
stars has been a controversial issue. That is, $M_{\rm trk}$ appears to be 
erroneously overestimated, which has been made known in comparison with the seismologically 
established mass ($M_{\rm seis}$) of higher precision. However, practical details of 
this problem (e.g., how much error is expected in which type of giants) are still unclear. 

In the meantime, another spectroscopic method for $M$ determination is conceptually possible,
which makes direct use of the surface gravity ($\log g$) along with $L$ and $T_{\rm eff}$
to derive the mass ($M_{gLT}$). Although this approach is simple and straightforward, it has 
barely been applied so far in practice, presumably because of the lack of reliability on $\log g$.

However, the author has come to develop some confidence on the surface gravity values
of red giants determined from Fe~{\sc i} and Fe~{\sc ii} lines in his previous studies,
because spectroscopic $\log g$ determined for giants in the {\it Kepler} field 
turned out to be consistent with seismic $\log g$ (T15, T16).

It was therefore decided to evaluate $M_{gLT}$ for those $\sim 100$ {\it Kepler} giants
(sample stars in T15 and T16), for which $M_{\rm seis}$  as well as the evolutionary status
(RG or RC1 or RC2) are already known, in order to examine the accuracy of $M_{gLT}$.
In addition, $M_{\rm trk}$ values were also determined for all the program stars by 
using the PARAM code (da Silva et al. 2006) for the purpose of clarifying the reliability
of $M_{\rm trk}$ under dispute. The necessary parameters ($T_{\rm eff}$, $\log g$, [Fe/H])
were taken from the previous work (mainly T15 and T16) and $L$ values were derived by using
the parallaxes (mainly from {\it Gaia} DR2). 

We could confirm a positive correlation between $M_{gLT}$ and $M_{\rm seis}$ (with a correlation
coefficient of +0.62), but $\log (M_{gLT}/M_{\rm seis})$ distribute over a wide range
between $\sim -0.3$ and $\sim +0.3$ (it average is +0.05 with $\sigma = 0.14$).
Therefore, the precision of $M_{gLT}$ by itself is rather insufficient for practical use, 
because errors on the order of several tens percent or more ($\sim 0.2$~dex) are quite common.
However, since this rather large uncertainty applies similarly to any star irrespective
of its type, studying the general trends of $M_{gLT}$ based on a large sample of stars 
may still be meaningful because random errors would be cancelled.

On the other hand, the trend of $M_{\rm trk}$ was found to be in marked contrast to the case 
of $M_{gLT}$, in the sense that it critically depends on the type of stars. Regarding RG 
and RC2 stars, the precision of $M_{\rm trk}$ is quite satisfactory  
because its error is typically within $\la$10--20\% as seen from the distribution 
of $\log (M_{\rm trk}/M_{\rm seis})$ (the mean is $-0.01$ and the standard deviation is 
0.06--0.08). However, for the case of RC1 stars 
(with comparatively lower mass of $0.8 \la M_{\rm seis}/M_{\odot} \la 1.5$) 
at the luminosity range of
$1.7 \la \log (L/L_{\odot}) \la 2.0$, $M_{\rm trk}$ is considerably higher than
$M_{\rm seis}$ with the difference amounting up to $\sim$100\% or more, which means that
$M_{\rm trk}$ is erroneously overestimated in this type of stars. 

Accordingly, since the evolutionary status of red giants is generally unknown in advance, 
the hybrid approach utilising both $M_{gLT}$ and $M_{\rm trk}$ in combination may  
be useful. $M_{\rm trk}$ may be preferably adopted in the first place, 
though utmost care should be taken for stars of $\log (L/L_{\odot}) \sim$~1.7--2.0, 
while $M_{gLT}$ may be secondarily employed to check if $M_{\rm trk}$ is not erroneously 
overestimated.

As an application of the potential merit of $M_{gLT}$ mentioned above, the age--metallicity 
relation of 322 red giants studied in T08 was revisited by using $M_{gLT}$ (instead of 
$M_{\rm trk,T08}$ which was later found to be rather overestimated) as the mass of each star. 
The new $age$ vs. [Fe/H] relation for 322 giants based on the revised mass values turned out 
to considerably mitigate the serious mismatch between giants and dwarfs reported in T08, 
indicating that statistically meaningful results may be obtained by using $M_{gLT}$.

The precision of $M_{gLT}$ is essentially determined by the accuracy of $\log g$.
However, since spectroscopic $\log g$ determination based on Fe~{\sc i} and 
Fe~{\sc ii} lines is a delicate procedure sensitively dependent upon the technical details
(e.g., selection of adopted lines), accomplishing a sufficient precision is not necessarily
easy. Therefore, in case of deriving $M_{gLT}$ by using Eq.(1), quantitative reliability 
of $\log g$ has to be confirmed in the first place.

\section*{Acknowledgments}

This research has been carried our by using the SIMBAD database,
operated by CDS, Strasbourg, France. 
For the purpose of mass determination in comparison with evolutionary tracks,
the open software PARAM (ver.1.3) was used via the web interface at the PARAM site 
maintained by Dr. Leo Girardi.

\section*{Data availability}

Regarding the open software tools (TGVIT, EXTINCT, PARAM), which are closely related 
to the contents of this article, their URLs are given in the footnotes of the main text.
The fundamental data (equivalent widths, abundances, atomic data) of Fe~{\sc i} and
Fe~{ii} lines, based on which the atmospheric parameters used in this study were
spectroscopically determined, are given in the electronic tables of the previous
papers of the author (referenced in Table~1).

\section*{Supplementary information}

The following online data (electronic data tables) are available as supplementary 
materials of this article, which gives the fundamental data and the final results 
of this investigation. 
\begin{itemize}
\item
{\bf readme.txt} 
\item
{\bf stellar\_parameters.dat} 
\item
{\bf PARAMresults.dat} 
\end{itemize}

\section*{Statements and declarations}

\subsection*{Funding}
The author declares that no funds, grants, or other support were received 
during the preparation of this manuscript.

\subsection*{Competing interests}
The author has no relevant financial or non-financial interests to disclose.

\subsection*{Author contributions}
This investigation has been conducted solely by the author.

\section*{Appendix: Effect of parallax correction on mass determination} 

Regarding the parallax data necessary for evaluating $L$ (luminosity) by using 
Eq.(3), we invoked {\it Gaia} DR2 values ($\pi_{Gaia}$) for all of the 103 giants 
in the {\it Kepler} field. After the release of these {\it Gaia} data, several studies 
argued the possibility of small zero-point offset to be applied to $\pi_{Gaia}$.
It is interesting to examine the impact of such small $\pi$ corrections in mass 
determination, given that evaluation of $M_{\rm trk}$ from intricate evolutionary 
tracks may be sensitive to a slight change in $L$.    

Zinn et al. (2019) reported based on asteroseismic data of red giants
in the {\it Kepler} field that zero-point offset exists in the 
{\it Gaia} DR2 parallaxes, which depends on {\it Gaia} astrometric 
pseudocolour ($\nu_{\rm eff}$) and {\it Gaia} $G$-band magnitude.
According to the formula derived by them, the values of this offset
($\pi_{\rm Zinn} - \pi_{Gaia}$) for each of our 103 {\it Kepler} giants 
($\nu_{\rm eff} \sim $~1.5--1.6, $G \sim$~6.7--12.1; cf. Fig.~8a) turned 
out $\sim$~0.05--0.06~m.a.s. as shown in Fig.~8b. 
This amount of slight increase in $\pi$ leads to a decrease in $\log L$
by $\la$~0.01--0.02~dex ($-0.4 \log (\pi_{\rm Zinn}/\pi_{Gaia})$).
Although such a small change in $\log L$ has a negligible effect in $M_{gLT}$
(only a few per cent at most), it might have an appreciable impact on $M_{\rm trk}$. 
We thus conducted a test determination of $M_{\rm trk}$ again based on PARAM
but adopting $\pi_{\rm Zinn}$ instead of $\pi_{Gaia}$ (other input parameters
were kept unchanged). The resulting mass values ($M_{\rm trk}^{\rm Z}$) are
compared with those of original $\pi_{Gaia}$-based $M_{\rm trk}$ in Fig.~8c,
where appreciable deviations (up to $\la$~20--30\%) are observed for some 
cases despite a reasonable correlation overall. A closer inspection revealed that 
such problematic stars are seen in the luminosity range of $\log (L/L_{\odot}) \sim$~1.5--2.0 
(cf. upper inset of Fig.~8d), which almost coincides with the region of large 
discrepancies between $M_{\rm trk}$ and $M_{\rm seis}$ (cf. Sect.~5.1). 
Accordingly, this effect of $M_{\rm trk}$ changes due to $\pi$ correction is eventually  
inconspicuous in the $\log (M_{\rm trk}/M_{\rm seis})$ vs. $\log L$ diagram (Fig.~8d), 
because stars showing appreciable differences (up to 
$\log (M_{\rm trk}^{\rm Z}/M_{\rm trk}) \la 0.1$~dex) just fall in the luminosity region 
where the dispersion is already large, as can be confirmed by comparing Fig.~8d with 
Fig.~5h (both appear almost similar). As such, we may state that the essential conclusion 
is not affected at all by using $\pi_{\rm Zinn}$ instead of $\pi_{Gaia}$.  

Throughout this study, the distance ($d$) was derived as the reciprocal of the parallax 
($1/\pi$) as usually done. However, Bailer-Jones et al. (2018) argued that distances 
obtained by simply inverting the {\it Gaia}~DR2 parallaxes may suffer appreciable errors.
Since this effect becomes progressively more significant with an increase in 
$\sigma_{\pi}/\pi$ (see Fig.~6 in Bailer-Jones et al., 2018; $\sigma_{\pi}$ is 
the error of $\pi$), it may be less significant in the present case where 
$\sigma_{\pi}/\pi$ is only several per cent at most. Anyway, we also checked how much 
correction is expected for our {\it Kepler}-field giants, because the revised distance 
data ($d_{\rm Bailer}$) are available in Bailer-Jones et al.'s (2018) catalogue.
Fig.~8e shows that the difference between $d_{\rm Bailer}$ and $d_{Gaia}$ 
($\equiv 1/\pi_{Gaia}$; adopted in this study) is almost zero at small $d$ but 
systematically grows towards larger $d$. 
Interestingly, in terms of the ``pseudo''-parallax (formally derived as 
$\pi^{*}_{\rm Bailer} \equiv d_{\rm Bailer}^{-1}$), the difference between these two
($\pi^{*}_{\rm Bailer} - \pi_{Gaia}$) is almost constant at $\simeq +0.03$~m.a.s. (Fig.~8f),
which means that this effect is equivalent to the zero-point offset of $\pi$
as long as our program stars are concerned. Considering that this offset is half as small as
that of Zinn et al. (2019) discussed above, we may similarly state that it does not have
any significant impact on our final conclusion. 

\newpage

\setcounter{table}{0}
\begin{table*}
\scriptsize
\caption{Basic data and the resulting parameters of 114 program stars.}
\begin{center}
\begin{tabular}{cccccccccccccc}\hline
Obj. name & $T_{\rm eff}$ & $\log g$ & [Fe/H] & Ref. & $V$ &
$A_{V}$ & $\pi$ & BC & $\log L$ & 
$M_{gLT}$  & $M_{\rm seis}$ & $M_{\rm trk}$ & Class \\
(1) & (2) & (3) & (4) & (5) & (6) & (7) & (8) & (9) & (10) & (11) & (12) &
(13) & (14) \\
\hline
KIC~01726211 & 4983& 2.49& $-$0.57& T15& 11.16& 0.44&  0.746& $-$0.27& 1.973& 1.91& 1.19& 1.61& RC1 \\
KIC~02013502 & 4913& 2.69& $-$0.02& T15& 11.52& 0.58&  0.680& $-$0.29& 1.977& 3.23& 1.94& 2.59& RC2 \\
KIC~02303367 & 4600& 2.39& +0.06& T15& 10.41& 0.30&  1.304& $-$0.43& 1.798& 1.40& 1.23& 1.49& RC1 \\
KIC~02424934 & 4792& 2.48& $-$0.18& T15& 10.55& 0.46&  1.004& $-$0.34& 1.996& 2.30& 1.36& 2.20& RC1 \\
KIC~02448225 & 4577& 2.37& +0.16& T15& 11.07& 0.39&  0.906& $-$0.44& 1.892& 1.69& 1.87& 1.97& RC2 \\
KIC~02573092 & 4688& 2.48& +0.00& T16& 11.70& 0.56&  0.692& $-$0.39& 1.920& 2.11& 1.43& 2.04& RC1 \\
KIC~02696732 & 4820& 2.90& $-$0.13& T16& 12.64& 0.45&  1.038& $-$0.33& 1.122& 0.79& 1.33& 1.15& RG  \\
KIC~02714397 & 4910& 2.56& $-$0.47& T15& 10.75& 0.43&  1.020& $-$0.29& 1.873& 1.89& 1.12& 1.35& RC1 \\
KIC~02988638 & 4912& 2.67& +0.06& T16& 12.15& 0.50&  0.597& $-$0.29& 1.804& 2.07& 2.31& 2.17& RC2 \\
KIC~03098045 & 4820& 2.34& $-$0.24& T16& 12.18& 0.49&  0.552& $-$0.33& 1.872& 1.22& 1.11& 1.53& RC1 \\
KIC~03217051 & 4590& 2.44& +0.21& T15& 11.87& 0.41&  0.714& $-$0.44& 1.783& 1.53& 1.22& 1.59& RC1 \\
KIC~03323943 & 4826& 2.55& $-$0.14& T16& 11.95& 0.51&  0.648& $-$0.32& 1.829& 1.79& 1.01& 1.57& RC1 \\
KIC~03425476 & 4780& 2.57& $-$0.03& T16& 12.56& 0.39&  0.604& $-$0.34& 1.606& 1.16& 1.37& 1.55& RC1 \\
KIC~03455760 & 4654& 2.68& $-$0.07& T15& 11.39& 0.43&  1.024& $-$0.40& 1.654& 1.86& 1.63& 1.18& RG  \\
KIC~03531478 & 5000& 3.20& $-$0.06& T16& 12.00& 0.29&  1.384& $-$0.26& 1.040& 1.13& 1.50& 1.28& RG  \\
KIC~03730953 & 4861& 2.55& $-$0.07& T15&  9.06& 0.19&  1.869& $-$0.31& 1.934& 2.21& 1.97& 2.44& RC2 \\
KIC~03744043 & 4946& 2.94& $-$0.35& T15&  9.81& 0.34&  2.476& $-$0.28& 1.437& 1.61& 1.31& 1.26& RG  \\
KIC~03748691 & 4762& 2.53& +0.11& T15& 12.25& 0.66&  0.601& $-$0.35& 1.847& 1.88& 1.37& 2.25& RC1 \\
KIC~03758458 & 5009& 2.71& +0.07& T15& 11.39& 0.55&  0.658& $-$0.26& 2.029& 3.53& 2.18& 2.75& RC2 \\
KIC~04036007 & 4916& 2.42& $-$0.36& T15& 11.72& 0.43&  0.607& $-$0.29& 1.933& 1.56& 1.38& 1.74& RC1 \\
KIC~04039306 & 4806& 2.45& $-$0.10& T16& 11.68& 0.51&  0.647& $-$0.33& 1.944& 1.88& 1.03& 2.28& RC1 \\
KIC~04044238 & 4519& 2.38& +0.20& T15&  7.80& 0.14&  4.131& $-$0.48& 1.796& 1.46& 1.06& 1.57& RC1 \\
KIC~04056266 & 5021& 2.67& $-$0.03& T16& 12.67& 0.49&  0.617& $-$0.25& 1.548& 1.05& 2.25& 1.96& RC2 \\
KIC~04243623 & 5005& 3.75& $-$0.31& T15& 11.40& 0.14&  2.830& $-$0.26& 0.599& 1.44& 0.99& 1.00& RG  \\
KIC~04243796 & 4620& 2.34& +0.11& T15& 10.86& 0.38&  1.000& $-$0.42& 1.877& 1.46& 1.26& 1.95& RC1 \\
KIC~04350501 & 4929& 3.19& $-$0.09& T16& 11.89& 0.50&  1.126& $-$0.28& 1.355& 2.41& 1.63& 1.40& RG  \\
KIC~04351319 & 4876& 3.32& +0.29& T15& 10.19& 0.15&  3.812& $-$0.30& 0.842& 1.04& 1.45& 1.27& RG  \\
KIC~04445711 & 4876& 2.50& $-$0.32& T15& 11.06& 0.36&  0.938& $-$0.31& 1.799& 1.43& 1.35& 1.35& RC1 \\
KIC~04448777 & 4805& 3.19& +0.10& T16& 11.87& 0.33&  1.579& $-$0.33& 1.018& 1.23& 1.13& 1.21& RG  \\
KIC~04570120 & 5035& 2.73& +0.05& T16& 11.46& 0.48&  0.613& $-$0.25& 2.032& 3.65& 2.44& 2.75& RC2 \\
KIC~04726049 & 5028& 3.25& $-$0.16& T16& 12.14& 0.48&  1.329& $-$0.25& 1.091& 1.39& 1.37& 1.26& RG  \\
KIC~04770846 & 4847& 2.60& +0.02& T15&  9.77& 0.33&  1.604& $-$0.32& 1.840& 2.02& 1.58& 2.28& RC1 \\
KIC~04902641 & 4986& 2.84& +0.03& T15& 11.14& 0.29&  1.085& $-$0.26& 1.594& 1.78& 2.56& 2.00& RC2 \\
KIC~04952717 & 4793& 3.13& +0.13& T15& 11.69& 0.42&  1.905& $-$0.34& 0.966& 0.96& 1.24& 1.20& RG  \\
KIC~05000307 & 5023& 2.64& $-$0.25& T15& 11.43& 0.46&  0.706& $-$0.25& 1.915& 2.28& 1.41& 2.27& RC1 \\
KIC~05033245 & 5049& 3.41& +0.11& T15& 11.24& 0.19&  2.389& $-$0.24& 0.823& 1.07& 1.41& 1.33& RG  \\
KIC~05088362 & 4760& 2.41& +0.03& T15& 11.38& 0.41&  0.622& $-$0.35& 2.064& 2.35& 2.22& 2.67& RC2 \\
KIC~05128171 & 4808& 2.54& +0.04& T15& 10.37& 0.42&  1.248& $-$0.33& 1.862& 1.91& 2.03& 2.32& RC2 \\
KIC~05266416 & 4767& 2.50& $-$0.09& T15& 10.80& 0.39&  1.001& $-$0.35& 1.875& 1.86& 1.51& 1.86& RC1 \\
KIC~05283798 & 4770& 2.53& +0.09& T16& 12.13& 0.50&  0.593& $-$0.35& 1.842& 1.84& 1.79& 2.14& RC2 \\
KIC~05307747 & 5030& 2.77& +0.01& T15&  8.67& 0.21&  2.731& $-$0.25& 1.745& 2.07& 2.91& 2.11& RC2 \\
KIC~05514974 & 4674& 2.26& $-$0.10& T16& 11.57& 0.43&  0.785& $-$0.39& 1.809& 0.99& 1.03& 1.42& RC1 \\
KIC~05530598 & 4599& 2.85& +0.37& T15&  8.93& 0.23&  3.986& $-$0.43& 1.393& 1.59& 1.68& 1.34& RG  \\
KIC~05598645 & 5052& 3.44& $-$0.17& T16& 12.10& 0.34&  1.406& $-$0.25& 1.000& 1.71& 1.14& 1.23& RG  \\
KIC~05611192 & 5064& 2.91& +0.02& T16& 12.12& 0.49&  0.611& $-$0.24& 1.771& 2.95& 2.43& 2.22& RC2 \\
KIC~05723165 & 5255& 3.67& $-$0.02& T15& 10.87& 0.26&  3.130& $-$0.19& 0.742& 1.37& 1.36& 1.35& RG  \\
KIC~05737655 & 5026& 2.45& $-$0.63& T15&  7.31& 0.17&  4.964& $-$0.26& 1.756& 1.02& 0.78& 1.00& RC1 \\
KIC~05795626 & 4923& 2.38& $-$0.72& T15&  9.39& 0.25&  1.530& $-$0.29& 1.993& 1.63& 1.21& 1.40& RC1 \\
KIC~05806522 & 4574& 2.60& +0.12& T15& 11.33& 0.40&  1.038& $-$0.45& 1.675& 1.74& 1.18& 1.25& RG  \\
KIC~05858034 & 4887& 2.45& $-$0.19& T16& 12.06& 0.44&  0.558& $-$0.30& 1.880& 1.52& 1.27& 1.89& RC1 \\
KIC~05866737 & 4874& 2.86& $-$0.26& T15& 10.91& 0.42&  1.198& $-$0.31& 1.670& 2.43& 1.52& 1.35& RG  \\
KIC~05990753 & 5010& 2.97& +0.19& T15& 11.04& 0.58&  1.010& $-$0.26& 1.810& 3.88& 2.70& 2.42& RC2 \\
KIC~06117517 & 4649& 2.94& +0.28& T15& 10.78& 0.40&  1.887& $-$0.41& 1.356& 1.72& 1.26& 1.31& RG  \\
KIC~06144777 & 4734& 3.02& +0.14& T15& 11.22& 0.31&  1.955& $-$0.37& 1.100& 1.06& 1.18& 1.22& RG  \\
KIC~06276948 & 4939& 2.84& +0.19& T15& 11.28& 0.34&  0.959& $-$0.28& 1.672& 2.21& 2.35& 2.06& RC2 \\
KIC~06531928 & 5156& 3.73& $-$0.57& T16& 10.97& 0.19&  3.211& $-$0.22& 0.662& 1.41& 0.95& 0.95& RG  \\
KIC~06579495 & 4772& 2.78& $-$0.02& T16& 12.20& 0.36&  0.902& $-$0.35& 1.392& 1.16& 1.22& 1.30& RG  \\
KIC~06665058 & 4750& 3.10& $-$0.07& T16& 11.47& 0.38&  1.282& $-$0.36& 1.390& 2.46& 1.18& 1.24& RG  \\
KIC~06690139 & 4979& 3.02& $-$0.14& T15& 11.86& 0.35&  0.898& $-$0.27& 1.498& 2.17& 1.56& 1.71& RG  \\
KIC~07205067 & 5064& 2.58& +0.03& T15& 10.80& 0.32&  0.725& $-$0.24& 2.085& 2.85& 3.49& 2.82& RC2 \\
KIC~07341231 & 5305& 3.65& $-$1.73& T16& 10.05& 0.14&  4.225& $-$0.22& 0.772& 1.35& 0.73& 0.80& RG  \\
\hline
\end{tabular}
\end{center}
\end{table*}

\setcounter{table}{0}
\begin{table*}
\scriptsize
\caption{(Continued.)}
\begin{center}
\begin{tabular}{cccccccccccccc}\hline
Obj. name & $T_{\rm eff}$ & $\log g$ & [Fe/H] & Ref. & $V$ &
$A_{V}$ & $\pi$ & BC & $\log L$ & 
$M_{gLT}$  & $M_{\rm seis}$ & $M_{\rm trk}$ & Class \\
(1) & (2) & (3) & (4) & (5) & (6) & (7) & (8) & (9) & (10) & (11) & (12) &
(13) & (14) \\
\hline
KIC~07581399 & 5070& 2.74& +0.01& T15& 11.54& 0.36&  0.586& $-$0.24& 1.988& 3.28& 3.13& 2.65& RC2 \\
KIC~07584122 & 4974& 3.29& $-$0.08& T16& 11.65& 0.29&  1.223& $-$0.27& 1.290& 2.52& 1.26& 1.41& RG  \\
KIC~07734065 & 4882& 2.20& $-$0.43& T16& 12.04& 0.43&  0.499& $-$0.31& 1.982& 1.09& 0.93& 1.65& RC1 \\
KIC~07799349 & 4969& 3.56& +0.28& T16&  9.67& 0.12&  5.434& $-$0.27& 0.717& 1.26& 1.31& 1.29& RG  \\
KIC~08378462 & 4995& 2.81& +0.06& T15& 11.40& 0.53&  0.878& $-$0.26& 1.769& 2.47& 2.47& 2.22& RC2 \\
KIC~08475025 & 4848& 2.88& $-$0.06& T16& 12.06& 0.34&  0.939& $-$0.32& 1.394& 1.38& 1.41& 1.36& RG  \\
KIC~08493735 & 5842& 3.62& +0.01& T16& 10.25& 0.17&  3.187& $-$0.08& 0.893& 1.13& 1.05& 1.47& RG  \\
KIC~08702606 & 5472& 3.66& $-$0.11& T15&  9.51& 0.04&  5.496& $-$0.14& 0.688& 1.01& 1.09& 1.26& RG  \\
KIC~08718745 & 4902& 3.20& $-$0.25& T15& 11.10& 0.30&  1.994& $-$0.30& 1.100& 1.40& 1.17& 1.13& RG  \\
KIC~08751420 & 5260& 3.63& $-$0.15& T16&  7.01& 0.05& 17.176& $-$0.19& 0.722& 1.19& 1.08& 1.27& RG  \\
KIC~08813946 & 4862& 2.64& +0.09& T15&  7.19& 0.14&  5.309& $-$0.31& 1.753& 1.79& 2.09& 2.02& RC1 \\
KIC~09145955 & 4943& 2.85& $-$0.34& T16& 10.05& 0.28&  2.384& $-$0.28& 1.350& 1.08& 1.39& 1.21& RG  \\
KIC~09173371 & 5064& 2.85& +0.00& T15&  9.59& 0.37&  2.111& $-$0.24& 1.657& 1.98& 2.32& 2.04& RC2 \\
KIC~09349632 & 4976& 2.75& +0.12& T16& 11.37& 0.38&  0.600& $-$0.27& 2.052& 4.19& 2.60& 2.81& RC2 \\
KIC~09583430 & 4854& 2.73& +0.21& T16& 12.03& 0.46&  0.733& $-$0.31& 1.666& 1.82& 2.62& 1.95& RC2 \\
KIC~09705687 & 5129& 2.81& $-$0.19& T15&  9.80& 0.28&  1.494& $-$0.22& 1.835& 2.58& 1.92& 2.21& RC2 \\
KIC~09812421 & 5140& 3.48& $-$0.21& T16& 10.32& 0.18&  3.327& $-$0.22& 0.887& 1.35& 1.27& 1.23& RG  \\
KIC~10323222 & 4525& 2.43& +0.04& T15&  7.02& 0.08&  7.366& $-$0.48& 1.578& 0.99& 1.55& 1.25& RG  \\
KIC~10382615 & 4890& 2.25& $-$0.49& T16& 12.02& 0.36&  0.528& $-$0.30& 1.911& 1.03& 1.00& 1.40& RC1 \\
KIC~10404994 & 4802& 2.63& $-$0.06& T15&  7.66& 0.12&  4.103& $-$0.33& 1.791& 2.01& 1.50& 1.65& RC1 \\
KIC~10426854 & 4968& 2.52& $-$0.30& T15& 10.31& 0.32&  1.399& $-$0.27& 1.724& 1.17& 1.78& 1.35& RC1 \\
KIC~10474071 & 4975& 2.65& +0.09& T16& 12.02& 0.50&  0.627& $-$0.27& 1.805& 1.89& 2.38& 2.34& RC2 \\
KIC~10600926 & 4879& 2.48& $-$0.20& T16& 11.88& 0.47&  0.627& $-$0.30& 1.865& 1.58& 0.82& 1.74& RC1 \\
KIC~10604460 & 4573& 2.37& +0.14& T16& 11.02& 0.41&  0.881& $-$0.45& 1.945& 1.91& 1.26& 2.08& RC1 \\
KIC~10709799 & 4522& 2.51& $-$0.04& T16& 11.29& 0.30&  1.086& $-$0.48& 1.623& 1.32& 1.29& 1.21& RG  \\
KIC~10716853 & 4874& 2.55& $-$0.08& T15&  7.04& 0.10&  5.386& $-$0.31& 1.783& 1.54& 1.75& 1.74& RC1 \\
KIC~10866415 & 4791& 2.82& $-$0.01& T16& 11.52& 0.44&  1.378& $-$0.34& 1.325& 1.08& 1.25& 1.29& RG  \\
KIC~11177749 & 4677& 2.24& $-$0.04& T16& 11.23& 0.33&  0.894& $-$0.39& 1.794& 0.92& 1.20& 1.44& RC1 \\
KIC~11251115 & 4837& 2.57& +0.07& T16&  8.09& 0.16&  3.088& $-$0.32& 1.877& 2.07& 2.45& 2.43& RC2 \\
KIC~11352756 & 4604& 2.29& $-$0.04& T16& 11.22& 0.47&  0.878& $-$0.43& 1.883& 1.34& 0.86& 1.57& RC1 \\
KIC~11401156 & 5053& 3.58& +0.10& T16& 10.01& 0.13&  4.807& $-$0.24& 0.682& 1.14& 1.09& 1.27& RG  \\
KIC~11444313 & 4757& 2.52& +0.00& T15& 11.65& 0.40&  0.678& $-$0.35& 1.880& 1.99& 1.39& 2.12& RC1 \\
KIC~11569659 & 4879& 2.48& $-$0.26& T15& 11.57& 0.58&  0.647& $-$0.31& 2.003& 2.18& 0.85& 2.28& RC1 \\
KIC~11618103 & 4922& 2.91& $-$0.17& T16&  7.93& 0.16&  6.176& $-$0.29& 1.326& 1.19& 1.41& 1.30& RG  \\
KIC~11657684 & 4951& 2.47& $-$0.12& T15& 12.00& 0.46&  0.548& $-$0.28& 1.914& 1.63& 1.27& 2.39& RC1 \\
KIC~11717120 & 5087& 3.72& $-$0.31& T16&  9.49& 0.12&  6.932& $-$0.24& 0.565& 1.17& 1.14& 1.00& RG  \\
KIC~11721438 & 4959& 2.87& +0.14& T16& 12.04& 0.48&  0.674& $-$0.27& 1.729& 2.66& 2.40& 2.14& RC2 \\
KIC~11802968 & 4962& 3.73& $-$0.06& T16& 11.22& 0.23&  3.327& $-$0.27& 0.569& 1.33& 0.81& 1.09& RG  \\
KIC~11819760 & 4824& 2.36& $-$0.18& T15& 11.27& 0.36&  0.692& $-$0.33& 1.986& 1.66& 1.25& 2.27& RC1 \\
KIC~12008680 & 4881& 2.55& $-$0.32& T16& 11.49& 0.45&  0.732& $-$0.30& 1.877& 1.91& 0.81& 1.56& RC1 \\
KIC~12070114 & 4698& 2.46& +0.05& T16& 11.05& 0.35&  0.817& $-$0.38& 1.949& 2.13& 1.78& 2.31& RC2 \\
KIC~12884274 & 4683& 2.44& +0.11& T15&  7.88& 0.20&  3.713& $-$0.39& 1.844& 1.62& 1.39& 2.10& RC1 \\
\hline
HD~107383    & 4841& 2.51& $-$0.28& T08&  4.74& 0.02& 10.710& $-$0.32& 2.081& 2.88& 2.48& 2.21& --- \\
HD~124897    & 4281& 1.72& $-$0.55& T09& $-$0.05& 0.00& 88.830& $-$0.64& 2.280& 1.21& 0.69& 1.04& --- \\
HD~062509    & 4904& 2.84& +0.06& T08&  1.14& 0.00& 96.540& $-$0.29& 1.592& 1.90& 2.31& 2.00& --- \\
HD~146791    & 4931& 2.69& $-$0.07& T08&  3.23& 0.00& 30.640& $-$0.28& 1.749& 1.88& 1.93& 1.75& --- \\
HD~168723    & 4972& 3.12& $-$0.18& T08&  3.25& 0.00& 53.930& $-$0.27& 1.245& 1.54& 1.82& 1.29& --- \\
HD~219615    & 4802& 2.25& $-$0.62& T08&  3.70& 0.04& 23.640& $-$0.34& 1.822& 0.90& 1.03& 1.12& --- \\
HD~185351    & 5006& 3.16& +0.00& T08&  5.17& 0.05& 24.225& $-$0.26& 1.187& 1.43& 2.03& 1.43& --- \\
HD~028307    & 4924& 2.63& +0.10& T08&  3.84& 0.00& 21.130& $-$0.29& 1.829& 1.98& 2.96& 2.39& --- \\
HD~076294    & 4844& 2.30& $-$0.11& T08&  3.10& 0.00& 19.510& $-$0.32& 2.206& 2.36& 3.15& 2.91& --- \\
Capella(P)   & 4943& 2.52& +0.10& T18&  0.92& 0.00& 76.200& $-$0.28& 1.880& 1.70& 2.57& 2.50& --- \\
Capella(S)   & 5694& 2.88& $-$0.08& T18&  0.75& 0.00& 76.200& $-$0.10& 1.878& 2.21& 2.49& 2.47& --- \\
\hline
\end{tabular}
\end{center}
(1) Object name. (2) Effective temperature (in K). (3) Logarithmic surface gravity (c.g.s. unit, 
in dex). (4) Logarithmic Fe abundance relative to the Sun (in dex).
(5) References for the spectroscopically determined parameters given in (2)--(4):
T08 $\cdots$ Takeda, Sato \& Murata (2008), T09 $\cdots$ Takeda et al. (2009),
T15 $\cdots$ Takeda \& Tajitsu (2015), T16 $\cdots$ Takeda et al. (2016), 
and T18 $\cdots$ Takeda, Hashimoto \& Honda (2018).  
(6) Apparent visual magnitude (in mag), where those of Capella's primary and secondary 
components were derived by adopting the flux ratio of $f_{\rm P}/f_{\rm S} = 0.85$ 
at 5500~\AA\ (cf. Fig.~1 in Torres et al., 2015). (7) Interstellar extinction (in mag).
(8) Trigonometric parallax (in milliarcsecond). (9) Bolometric correction (in mag).
(10) Logarithmic bolometric luminosity (in unit of $L_{\odot}$).
(11) Stellar mass derived from $g$, $L$, and $T_{\rm eff}$ by using Eq.(1).
(12) Reference mass values determined by the seismic technique (except for those 
of the binary system Capella, which were derived from their orbital motions).
(13) Stellar mass derived from the positions on the HR diagram in comparison with
theoretical evolutionary tracks by using PARAM.
(14) Evolutionary class of each {\it Kepler} giant seismologically determined 
by Mosser et al. (2012) (RG: red giant, RC1: 1st clump giant, RC2: 2nd clump giant). 
\end{table*}

\setcounter{figure}{0}
\begin{figure*}
\begin{center}
\begin{minipage}{80mm}
\includegraphics[width=8.0cm]{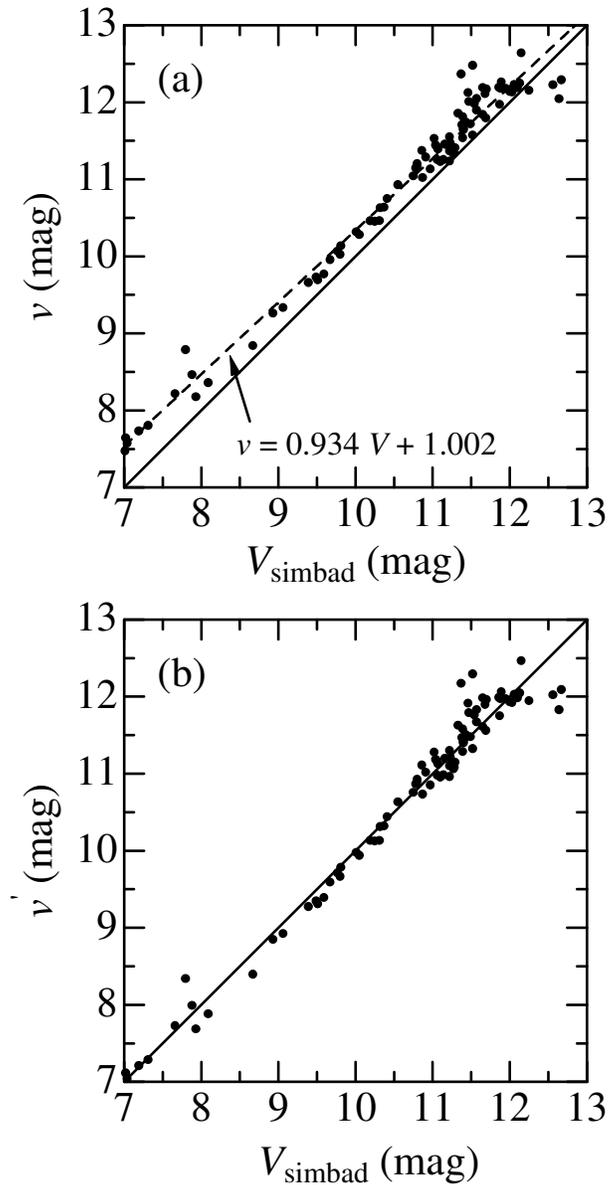}
\caption{(a) The visual-band magnitudes ($v$), which were tentatively estimated 
for the {\it Kepler} giants based on the $r$-band magnitudes given in {\it Kepler} 
Input Catalogue (Brown et al., 2011) by using the theoretical $V-R$ indices of 
Giraldi et al.'s (2000) stellar evolution calculations, 
plotted against the actual $V$ magnitudes taken from the SIMBAD database.
The linear-regression relation ($v = 0.934 V + 1.002$) is also depicted by the dashed line.
(b) The corrected $v'$ values are plotted against $V$, where $v' \equiv (v - 1.002)/0.934$. 
}
\label{fig1}
\end{minipage}
\end{center}
\end{figure*}

\setcounter{figure}{1}
\begin{figure*}
\begin{center}
\begin{minipage}{120mm}
\includegraphics[width=12.0cm]{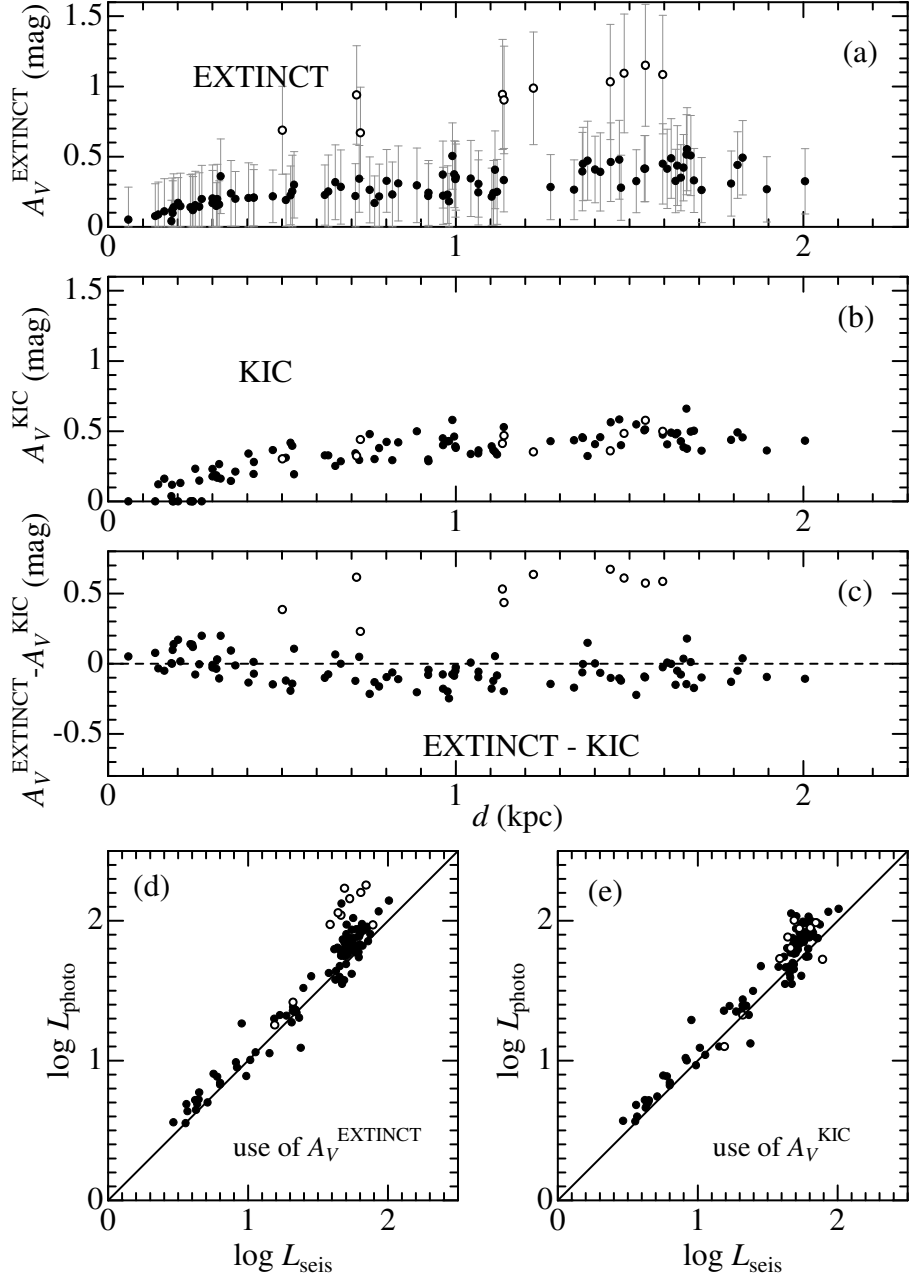}
\caption{
(a) Visual extinction values ($A_{V}^{\rm EXTINCT}$), which were evaluated with the help
of the EXTINCT code (Hakkila et al., 1997) for the 103 {\it Kepler} giants, plotted against 
the distance, where error bars indicate the total extinction errors given by EXTINCT.
(b) $A_{V}^{\rm KIC}$ values of {\it Kepler} giants, which are given in the {\it Kepler} 
Input Catalogue (Brown et al., 2011), plotted against the distance. 
(c) Difference between $A_{V}^{\rm EXTINCT}$ and $A_{V}^{\rm KIC}$, plotted against the distance.
(d) Luminosities of {\it Kepler} giants ($L_{\rm photo}$) photometrically derived
based on Eq.(2) with $A_{V}^{\rm EXTINCT}$ are compared with those ($L_{\rm seis}$) 
derived in T15 and T16 by using $R_{\rm seis}$ (seismic radius).
(e) $L_{\rm photo}$ values of {\it Kepler} giants obtained by using $A_{V}^{\rm KIC}$ 
are plotted against $L_{\rm seis}$.
In each panel, 10 stars showing appreciable deviations of $A_{V}^{\rm EXTINCT}$ from the main 
trend in panel~(a) are plotted by open symbols, in order to distinguish them from others.  
}
\label{fig2}
\end{minipage}
\end{center}
\end{figure*}

\setcounter{figure}{2}
\begin{figure*}
\begin{center}
\begin{minipage}{100mm}
\includegraphics[width=10.0cm]{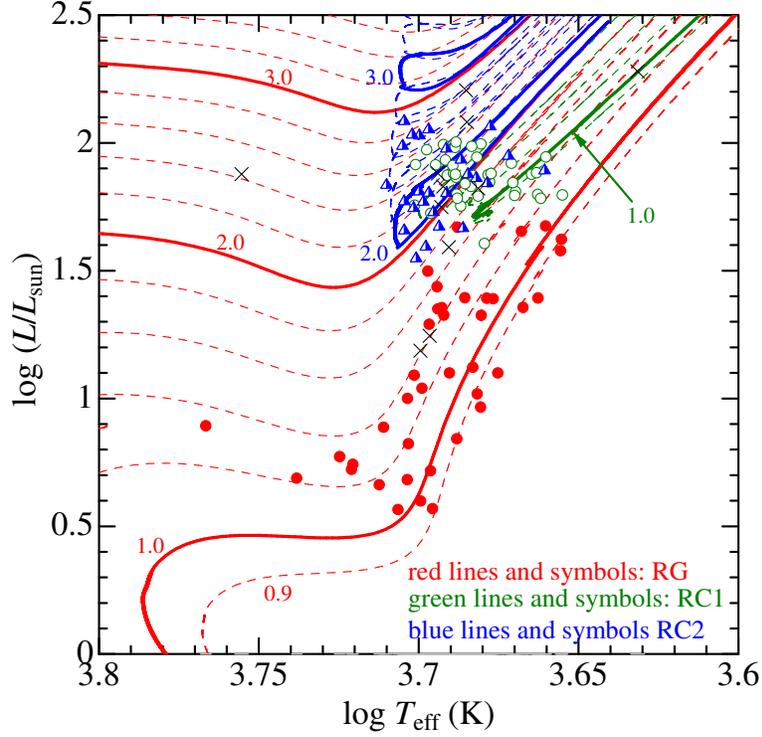}
\caption{
The 114 program stars plotted on the $\log T_{\rm eff}$--$\log L$ diagram,
where the PARSEC tracks computed for $z = 0.01$ (slightly metal-deficient 
case by $\sim 0.2$~dex lower than the solar metallicity) and various $M$ values (0.9, 1.0,
1.2, 1.4, 1.6, 2.0, 2.2, 2.4, 2.6, 2.8, 3.0, 3.2, 3.4~$M_{\odot}$, thick solid lines for 
the cases of integer masses, otherwise dashed lines) are overplotted for comparison. 
While 11 nearby stars are indicated simply by black crosses, the symbols for 103 {\it Kepler} giants 
are discriminated according to the evolutionary status: shell-H-burning phase before He ignition (RG) 
--- red filled circles, $M \le 1.8$~$M_{\odot}$ stars at the He-burning phase (RC1) 
--- green open circles, and $M > 1.8$~$M_{\odot}$ stars at the He-burning phase (RC2) 
--- blue half-filled triangles.
Similarly, the evolutionary tracks are depicted in the relevant colours (red or green or blue)
corresponding to their evolutionary stages. 
}
\label{fig3}
\end{minipage}
\end{center}
\end{figure*}

\setcounter{figure}{3}
\begin{figure*}
\begin{center}
\begin{minipage}{80mm}
\includegraphics[width=8.0cm]{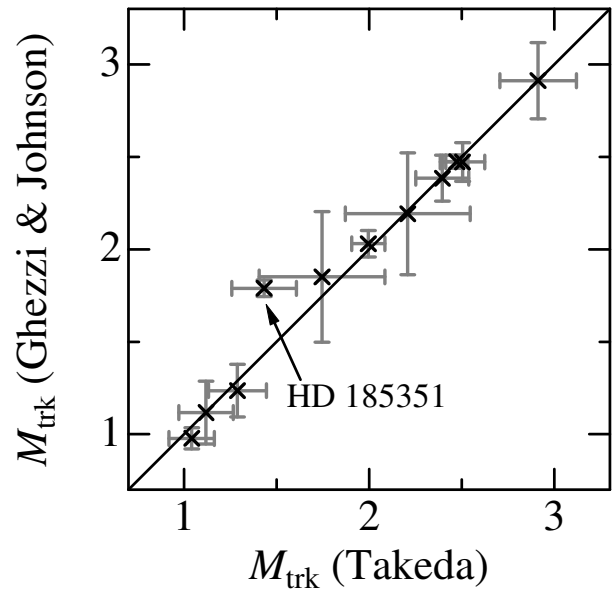}
\caption{
Comparison of the masses derived by using PARAM in this study (abscissa) with those 
of Ghezzi \& Johnson (2015) (ordinate) for 11 nearby giant stars in common.
}
\label{fig3}
\end{minipage}
\end{center}
\end{figure*}

\setcounter{figure}{4}
\begin{figure*}
\begin{center}
\begin{minipage}{120mm}
\includegraphics[width=12.0cm]{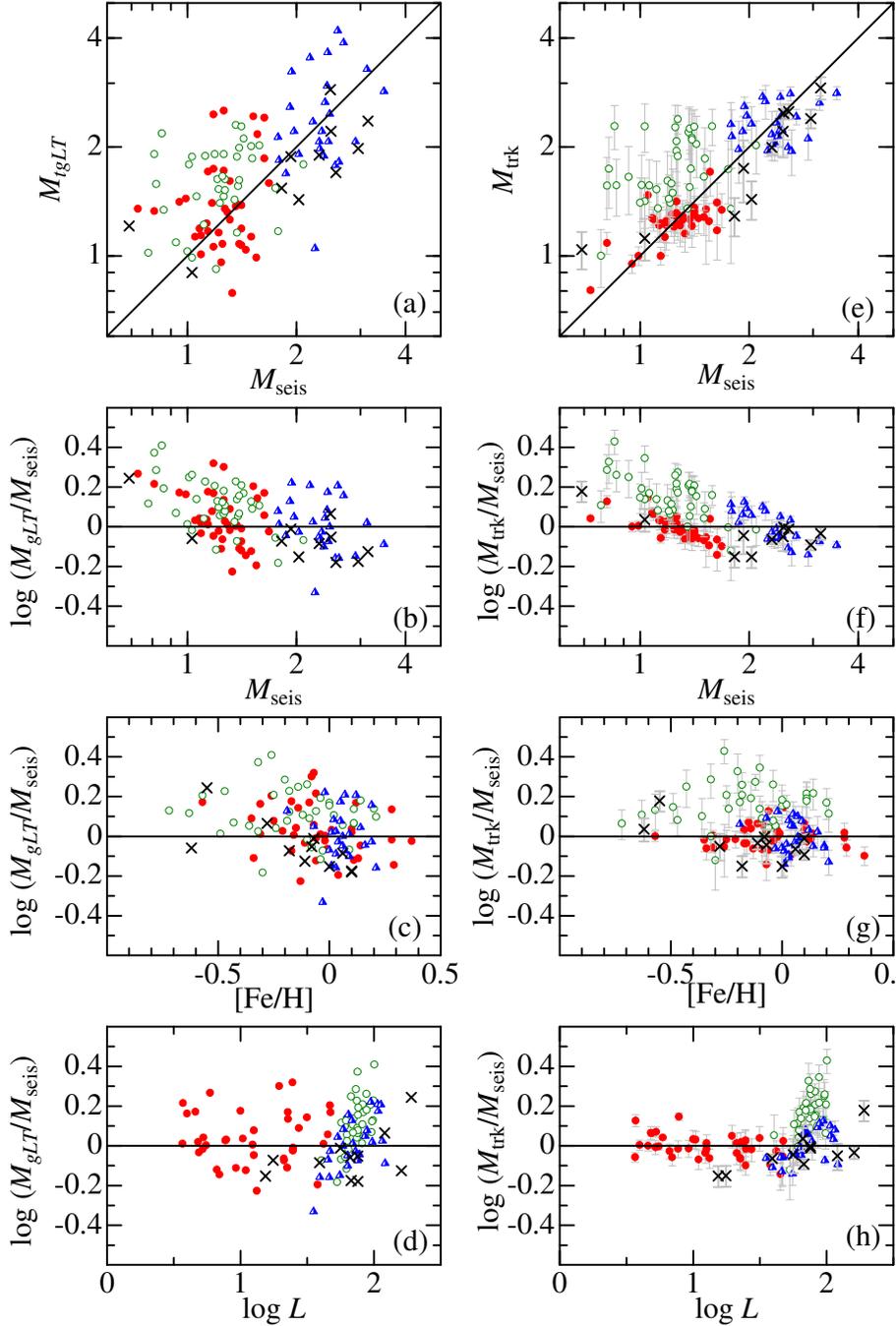}
\caption{
Left-hand panels show how the $M_{gLT}$ values (mass derived from $g$, $L$, and $T_{\rm eff}$) 
of 114 program stars are compared with $M_{\rm seis}$ (reference mass established 
by the seismic technique). Comparison between $M_{gLT}$ and $M_{\rm seis}$ is depicted
in panel (a), while panels (b)--(d) display how $\log (M_{gLT}/M_{\rm seis})$ 
correlates with $M_{\rm seis}$, [Fe/H], and $\log L$, respectively.
Right-hand panels (e)--(h) similarly illustrate the behaviours of $M_{\rm trk}$ 
(mass derived from the position on the HR diagram by using PARAM) 
against $M_{\rm seis}$. The meanings of the symbols are the same as in Fig.~3.
Note that the axis for the mass is in the logarithmic scale. 
}
\label{fig4}
\end{minipage}
\end{center}
\end{figure*}

\setcounter{figure}{5}
\begin{figure*}
\begin{center}
\begin{minipage}{120mm}
\includegraphics[width=12.0cm]{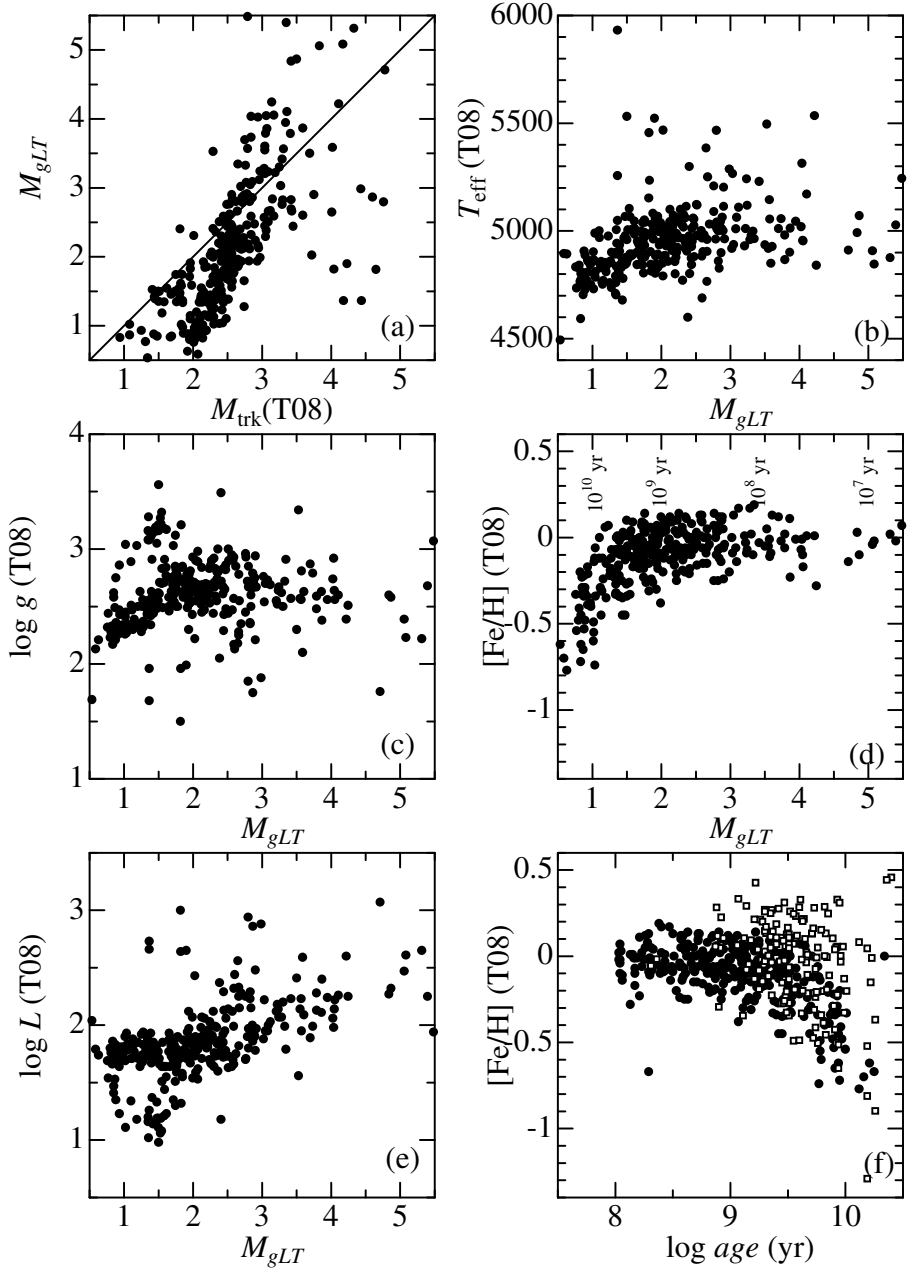}
\caption{Trends of the revised mass values ($M_{gLT}$) of 322 red giants studied in T08, 
which were derived from their $\log g$, $\log L$, and $T_{\rm eff}$ by using Eq.(1). 
Panel (a) shows the relation between $M_{gLT}$ and $M_{\rm trk}$(T08) (derived in T08  
based on theoretical evolutionary tracks), 
while panels (b)--(e) illustrate how $T_{\rm eff}$, $\log g$, [Fe/H], and $\log L$ 
correlate with $M_{gLT}$, respectively. The revised [Fe/H] vs. {\it age} plots
({\it age} was derived from $M_{gLT}$; cf. Sect.~5.2) are depicted in panel (f), 
where the age--metallicity relations for 160 FGK dwarfs (+subgiants) derived in 
Takeda (2007) are also overplotted in open squares for comparison.
}
\label{fig5}
\end{minipage}
\end{center}
\end{figure*}

\setcounter{figure}{6}
\begin{figure*}
\begin{center}
\begin{minipage}{120mm}
\includegraphics[width=12.0cm]{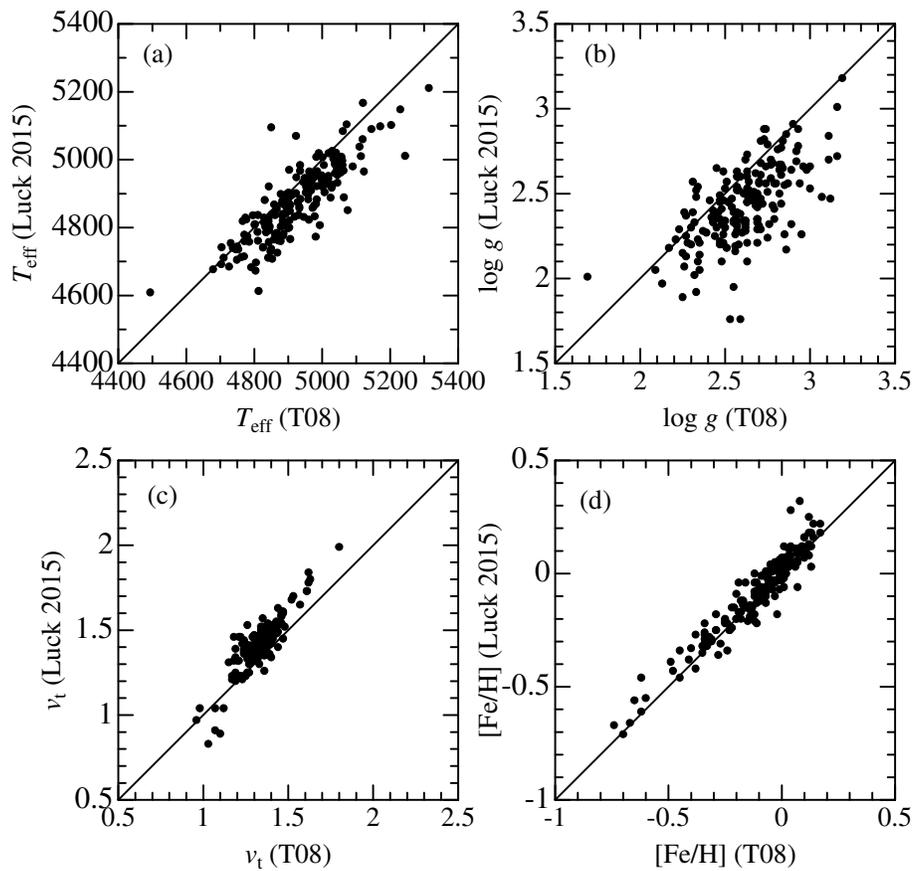}
\caption{
Comparison of the atmospheric parameters spectroscopically determined in T08
with those derived by Luck (2015) for 189 stars in common: 
(a) $T_{\rm eff}$, (b) $\log g$, (c) $v_{\rm t}$ (microturbulence), and (d) [Fe/H].
}
\label{fig6}
\end{minipage}
\end{center}
\end{figure*}

\setcounter{figure}{7}
\begin{figure*}
\begin{center}
\begin{minipage}{120mm}
\includegraphics[width=12.0cm]{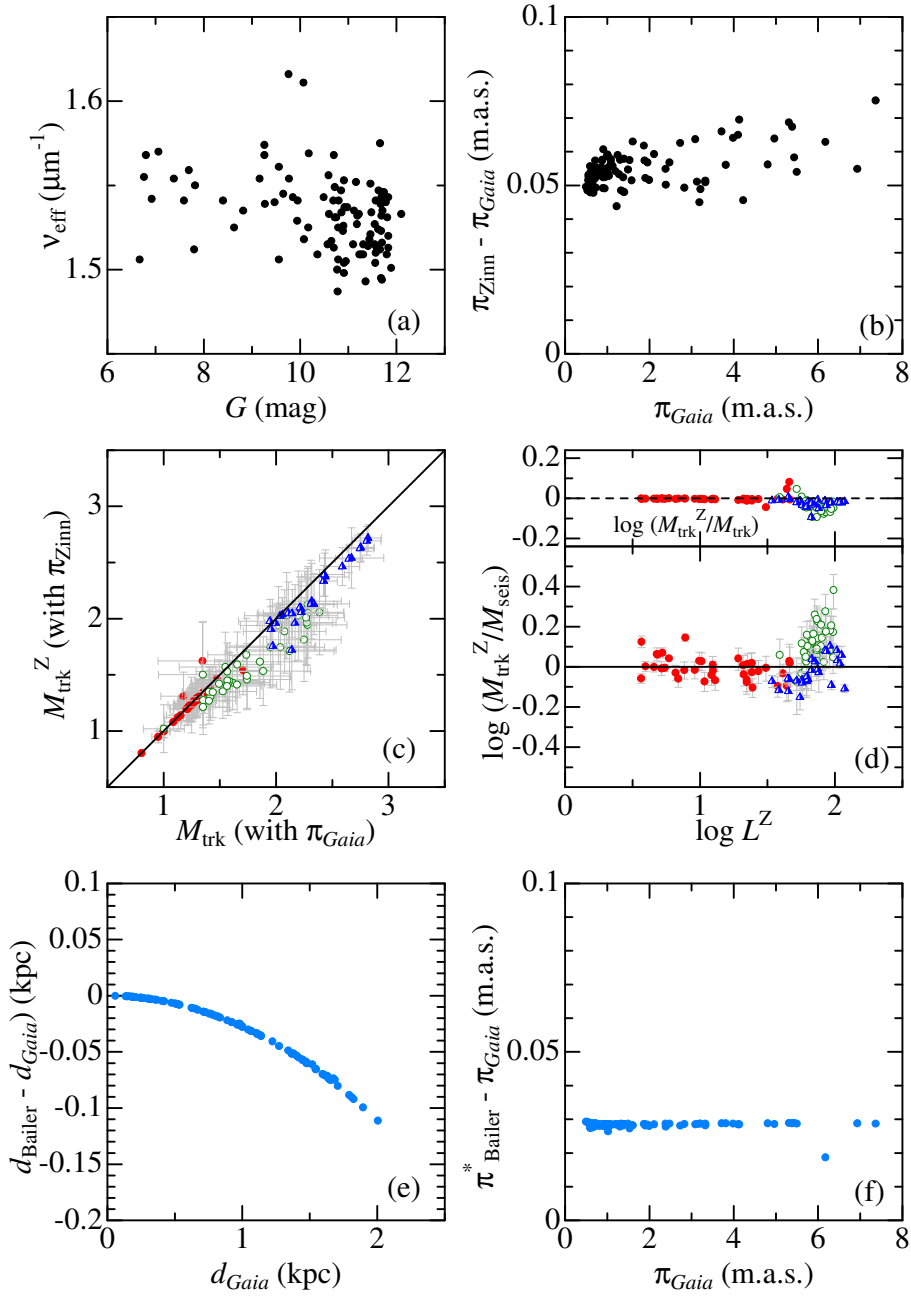}
\caption{
(a) Correlation between $G$ (magnitude in the $G$-band) and $\nu_{\rm eff}$ (astrometric 
pseudocolour) for the 103 {\it Kepler} giants (taken from {\it Gaia} DR2 catalogue).
(b) Zero-point offset corrections ($\pi_{\rm Zinn} - \pi_{Gaia}$), which were evaluated 
from $G$ and $\nu_{\rm eff}$ by using Zinn et al.'s (2016) relation, plotted against $\pi_{Gaia}$. 
(c) Correlation between $M_{\rm trk}^{\rm Z}$ (derived by using PARAM but with $\pi_{\rm Zinn}$) 
and $\pi_{Gaia}$-based $M_{\rm trk}$ derived in Sect.~4, where the meanings of the symbols are
the same as in Fig.~3.
(d) The logarithmic ratios of $\log (M_{\rm trk}^{Z}/M_{\rm seis})$ plotted against $\log L^{\rm Z}$ 
($\pi_{\rm Zinn}$-based luminosity), which is arranged similarly to Fig.~5h for the sake
of comparison. Note that $\log L^{\rm Z}$-dependence of $\log (M_{\rm trk}^{\rm Z}/M_{\rm trk})$ 
is also depicted in the upper inset of this panel.
(e) Distance differences for 103 {\it Kepler} giants between Bailer-Jones et al.'s (2018) values
($d_{\rm Bailer}$) and those derived from {\it Gaia} DR2 parallaxes ($d_{Gaia} \equiv \pi_{Gaia}^{-1}$), 
plotted against $d_{Gaia}$.
(f) Difference between the ``pseudo''-parallax (formally derived as 
$\pi^{*}_{\rm Bailer} \equiv d_{\rm Bailer}^{-1}$) and $\pi_{Gaia}$, plotted against $\pi_{Gaia}$.
}
\label{fig8}
\end{minipage}
\end{center}
\end{figure*}

\end{document}